\documentclass[11pt,twoside]{article}
\usepackage{./asp2014}

\aspSuppressVolSlug
\resetcounters

\begin{document}

\title{Seismology of an Ensemble of ZZ Ceti Stars}
\author{ J.~C.~Clemens$^1$, P.~C.~O'Brien$^1$, Bart.~H.~Dunlap$^1$, J.~J.~Hermes$^{1, 2}$
\affil{$^1$University of North Carolina at Chapel Hill, Chapel Hill, NC, USA; \email{clemens@physics.unc.edu}}
\affil{$^2$Hubble Fellow}}

\paperauthor{ J.~C.~Clemens}{clemens@physics.unc.edu}{}{University of North Carolina at Chapel Hill}{Department of Physics and Astronomy}{Chapel Hill}{NC}{27599}{USA}
\paperauthor{P.~C.~O'Brien}{pcobrien@live.unc.edu}{}{University of North Carolina at Chapel Hill}{Department of Physics and Astronomy}{Chapel Hill}{NC}{27599}{USA}
\paperauthor{ B.~H.~Dunlap}{bhdunlap@physics.unc.edu}{}{University of North Carolina at Chapel Hill}{Department of Physics and Astronomy}{Chapel Hill}{NC}{27599}{USA}
\paperauthor{ J.~J.~Hermes}{jjhermes@email.unc.edu}{}{University of North Carolina at Chapel Hill}{Department of Physics and Astronomy}{Chapel Hill}{NC}{27599}{USA}

\begin{abstract}
We combine all the reliably-measured eigenperiods for hot, short-period ZZ Ceti stars onto one diagram and show that it has the features expected from evolutionary and pulsation theory.  To make a more detailed comparison with theory we concentrate on a subset of 16 stars for which rotational splitting or other evidence gives clues to the spherical harmonic index ($\ell$) of the modes.  The suspected $\ell=1$ periods in this subset of stars form a pattern of consecutive radial overtones that allow us to conduct ensemble seismology using published theoretical model grids.  We find that the best-matching models have hydrogen layer masses most consistent with the canonically thick limit calculated from nuclear burning.   We also find that the evolutionary models with masses and temperatures from spectroscopic fits cannot correctly reproduce the periods of the $k =$ 1 to 4 mode groups in these stars, and speculate that the mass of the helium layer in the models is too large.
\end{abstract}

\section{Introduction}
Long after the recognition that ZZ Ceti pulsations might be used to understand the internal structures of white dwarfs, these stars remain enigmatic.   For most individual ZZ Ceti stars with stable, resolved pulsation spectra, the number of periods is small and the radial order ($k$) and spherical harmonic degree ($\ell$) unknown.  As a result, any comparison between the eigenmodes of pulsational models and observed periods is underconstrained.  Thus, even when a few observed periods in a star are well-matched by a few eigenperiods from a model, we cannot be certain that this match implies the model has a structure resembling that of the actual star.   This shortcoming in ZZ Ceti seismology has been addressed in a variety of ways: by using fully evolutionary models with diffusion to reduce the number of free parameters that must be fitted \citep{romero2012}; by improving the spectroscopic determinations of mass and gravity \citep{gianninas2011, tremblay2013}; and by long-timebase or high-sensitivity observing campaigns to detect more modes and resolve them into multiplets that can constrain $\ell$ if not $k$ \citep{giammichele2015}.  

\citet{clemens1993,clemens1994} noticed that the short period modes from hot ZZ Ceti stars compiled into one diagram defined a pattern with distinct period groupings separated by gaps, and suggested that studying the structure in this ensemble diagram might help to resolve the degeneracy in the fitting process.   One reason for this is that geometric cancellation should result in the largest modes being mostly $\ell=1$, so uncertainties arising from individual mode identification are reduced in the ensemble diagram.  Another is that the mean periods of the groups formed by the largest modes depend on the average mass and temperature of the ensemble, which are more robustly known from spectroscopy than those for any star alone, reducing the number of free parameters.     

For stars with a narrow distribution in mass, as is known to be true for white dwarfs, a pattern like that found by \citet{clemens1994} is expected from pulsation theory.  The left hand panel of Figure \ref{fig:clemens_j_fig1}, reproduced from \citet{romero2012} illustrates this point.  The figure shows low-order $\ell=1$ eigenmode periods for evolutionary models near the mean of all white dwarf masses.  For a wide variety of hydrogen layer masses ($M_{\rm H}$), which is treated as a free parameter, the periods of the low order modes will accumulate near one of the solid yellow horizontal lines in the diagram.  Although these theoretical periods will be different for models with mass not equal to the value in this plot (0.593 $M_{\odot}$), the narrowness of the observed field white dwarf mass distribution \citep{tremblay2016} allows the $\ell=1$ period groups to remain distinct for the lowest radial overtones.

\articlefigure[width=0.95\textwidth]{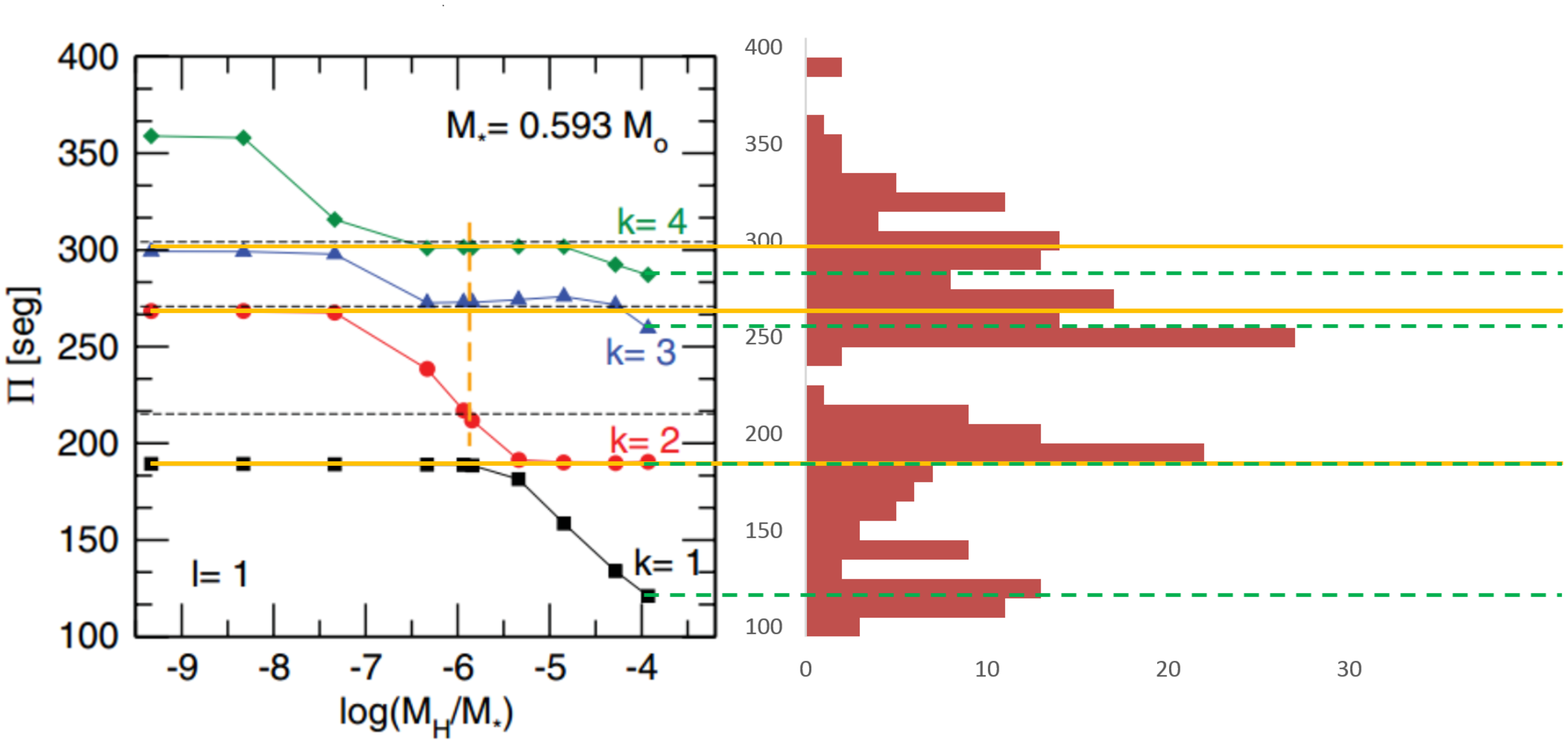}{fig:clemens_j_fig1}{The left-hand panel, reproduced from \citet{romero2012}, shows a sequence of pulsation periods derived from their evolutionary models with $M=0.593 M_\odot$.  Given random choices for $M_{\rm H}$, $\ell=1$ mode periods will accumulate near the solid yellow lines, while choosing canonically thick $M_{\rm H}$ will give periods near the dotted green lines.  The histogram on the right shows the distribution of mode periods for all 239 modes measured in 75 hot ZZ Ceti stars.}

\section{Updating the Hot ZZ Ceti Period Ensemble}

\citet{clemens1994} collected the observed periods for 11 hot ZZ Ceti stars and found suggestions of an emergent pattern, but small-number statistics and lack of certain $\ell$ identification limited the application of the ensemble technique.  Now, 22 years later, we have repeated this exercise for 75 known hot ZZ Cetis, most of them still without secure mode identification.  As the top panel of Figure \ref{fig:clemens_j_fig2} shows, the evidence for grouping is still apparent.  The gap in periods between 220 and 250 s is particularly striking, and aligns with the region between predicted mode periods in the models of Figure \ref{fig:clemens_j_fig1}.  The existence of this gap and other features argues against allowing the hydrogen layer mass to vary freely, because $\ell=1$, $k=2$ models in the crossing region near $M_{\rm H} = 10^{-6}M_\star$ would populate the gap, conflicting with the observations.  Likewise, the appearance of a grouping of periods in the 100-120 s range suggests that a significant fraction of the stars have the thickest possible $M_{\rm H}$ allowed by nuclear burning, which is around $10^{-4} M_\star$ for this model.  To illustrate these points we have converted the period distribution in the top panel of Figure \ref{fig:clemens_j_fig2} into a histogram and added it to the right of the Romero diagram in Figure \ref{fig:clemens_j_fig1}.  This ensemble of observed periods still suffers from the drawback that we do not know which modes are $\ell=1$, so the observed distribution likely combines modes of different $\ell$, while the theoretical diagram has only periods for $\ell=1$ modes.  Likewise, we do not know the azimuthal degree $m$ of the modes, but for periods below 400 s and typical rotation periods of 1 d, the ambiguities from $m$ identification are typically less than 1 s.

\articlefigure[width=0.95\textwidth]{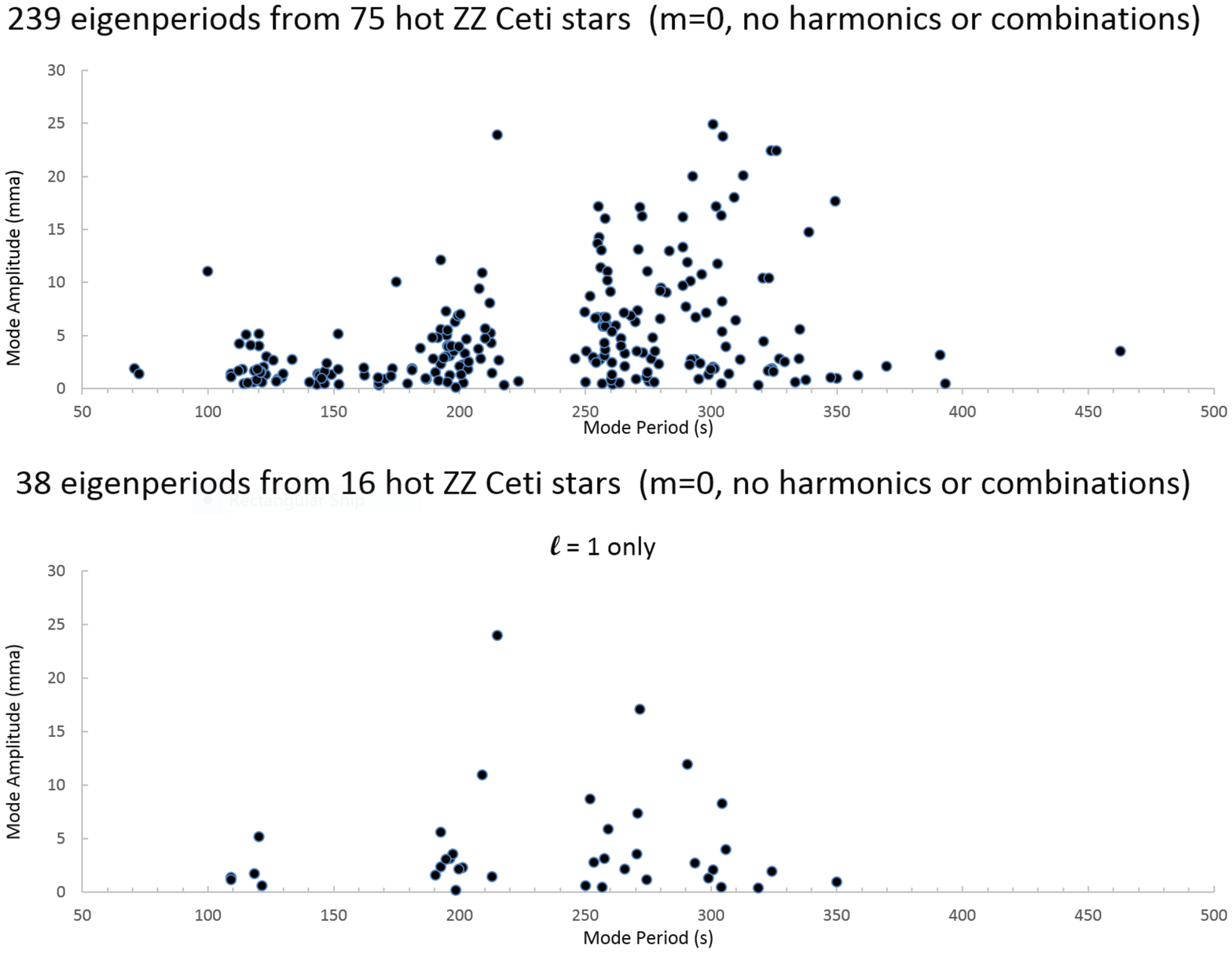}{fig:clemens_j_fig2}{The top panel shows all reliably-measured eigenperiods for the 75 known short-period (hot) ZZ Cetis.  The bottom panel are the likely $\ell=1$ periods for 16 stars in which common rotational splittings or other methods allow $\ell$ identification. These appear to form a sequence of continuous radial overtone, k, from 1 to 4.  All combination frequencies have been ignored, and multiplets are plotted once at the central period of the multiplet (assumed $m=0$).}

Fortunately, it is now possible to improve this diagram by restricting ourselves to a subset of stars for which we have independent knowledge of the sperical degree of the eigenperiods observed.  The number of stars for which this is possible has recently increased owing to discovery and short cadence observations of new ZZ Ceti stars by the {\it Kepler} and {\it K2} missions. These new ZZ Cetis will be the subject of a series of forthcoming papers by \citet{hermes2016}.  Figure \ref{fig:clemens_j_fig3} shows the Fourier Transform of {\it K2} data accumulated over 78.7 days at 60 s cadence for one such star.  In addition, improved ground-based observations of GD 165 and R 548 (ZZ Ceti) \citep{giammichele2015}, GD 66 and G238-53 \citep{mullally2008} , LP133-144 \citep{bognar2016}, and other well-known stars have added to our knowledge of their multiplet structure. The subset of hot ZZ Ceti stars with $\ell$ identification we consider in this paper includes eight new stars from {\it Kepler} and eight previously known examples.  The lower panel of Figure \ref{fig:clemens_j_fig2} shows the periods of those modes in these stars which we identify as $\ell=1$  on the basis of rotational splitting or, in the case of G117-B15a, arguments from multicolor photometry \citep{robinson1995} and combination mode amplitudes \citep{yeates2005}.  

\articlefigure[width=0.95\textwidth]{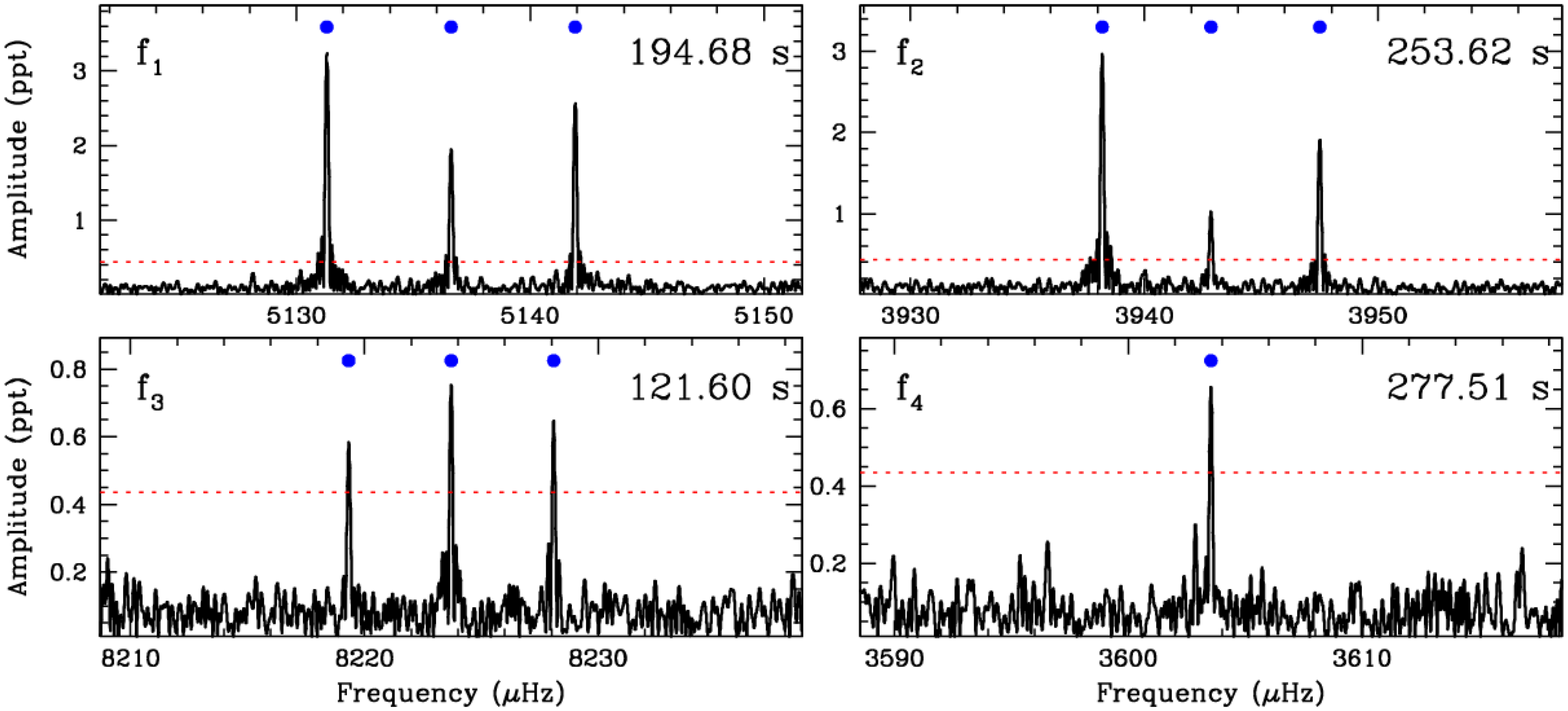}{fig:clemens_j_fig3}{Typical multiplet structure from {\it Kepler} short cadence observations of a hot ZZ Ceti star.  We identify the modes at 121.6, 194.7, and 253.6 s in this star as $\ell = 1$, and discard 277.5 s because it cannot be securely identified.}

\section{Monte Carlo Simulations using the Romero et.al Period Grid}

In order to compare these measured $\ell =1$ eigenperiods with the models, we have converted the lower panel of Figure \ref{fig:clemens_j_fig2} into a histogram that appears in the top panel of Figure \ref{fig:clemens_j_fig4}.  We interperet this histogram as a sequence of consecutive $k$ from 1 to 4.  To compare these periods to the seismological prediction of evolutionary models, we used the published grid of $\ell=1$ periods from \citet{romero2012} to conduct Monte Carlo simulations of the observed sample.  Using the measured spectroscopic masses and temperatures for the stars, with suitable Gaussian errors applied, we drew 50 simulated period spectra for the sample of 16 stars (1900 periods in all), keeping only those $k$ for each star that match the ones observed in its period spectrum.   

We have plotted these simulations for a number of assumptions in the four lower panels of Figure \ref{fig:clemens_j_fig4}.  The second and third panels show the simulated periods drawn from the Romero models under the assumption that all stars have the canonically thick $M_{\rm H}$.   The temperatures and gravities used in the second panel are from the 1D models of \citet{gianninas2011} while the third panel shows the values corrected for the three-dimensional dependence of convection \citep{tremblay2013}.    In their overall appearance, these two simulations most resemble the data.  They have a group at 100-120 s, and clean gaps between the lower $k$ modes.   However, the periods of the $k>1$ mode groups do not match the data; the spacing between modes of consecutive $k$ in the models is systematically too small.  We experimented with uniformly decreasing the masses in the models to spread out the overtones, and found that a roughly 10\% decrease would be required, and this would still not improve the match to the $k=2$ group.  A systematic mass error of 10\% in all the stars is implausible, but a more promising approach is to alter the mass of the helium layer $M_{\rm He}$.  We have used quasi-evolutionary models to experiment with thinner layers and this approach shows promise.  We will pursue this approach further in a forthcoming paper.

Finally, we have also calculated simulations in which $M_{\rm H}$ is allowed to vary from the canonical limit.  These do not match the data as well inasmuch as they lack a clear group at 100-120 seconds, and the gaps between consecutive $k$ are not clean.  This comports with the expectations we developed based on Figure \ref{fig:clemens_j_fig1};  varying $M_{\rm H}$ allows stars to have periods in between the yellow accumulation lines on the diagram and does not select the correct  fraction of stars with modes near 100 s.   The fourth panel shows models in which $M_{\rm H}$ was treated as a random variable, while the bottom panel uses the distribution of $M_{\rm H}$ found by \citep{romero2012} using individual seismological fits to an ensemble of stars with mostly unknown values of $\ell$ and $k$.  

\section{Conclusions}

Examining the ensemble of hot ZZ Ceti pulsators with known $\ell$ shows a pattern of consecutive groups that we have identified with the lowest $k$ non-radial $g$-mode pulsations.  Analyzing this pattern offers a promising way to constrain seismological model parameters like $M_{\rm H}$ and $M_{\rm He}$.  The 16 stars in our sample suggest that most hot ZZ Cetis have $M_{\rm H}$ values at or near the canonical limit and that their helium layers are thinner than those calculated  by evolutionary models.  Detailed model fits to individual stars informed by these results should allow us to extract believable asteroseismological measurements of ZZ Ceti interior properties.

\articlefigure[width=0.95\textwidth]{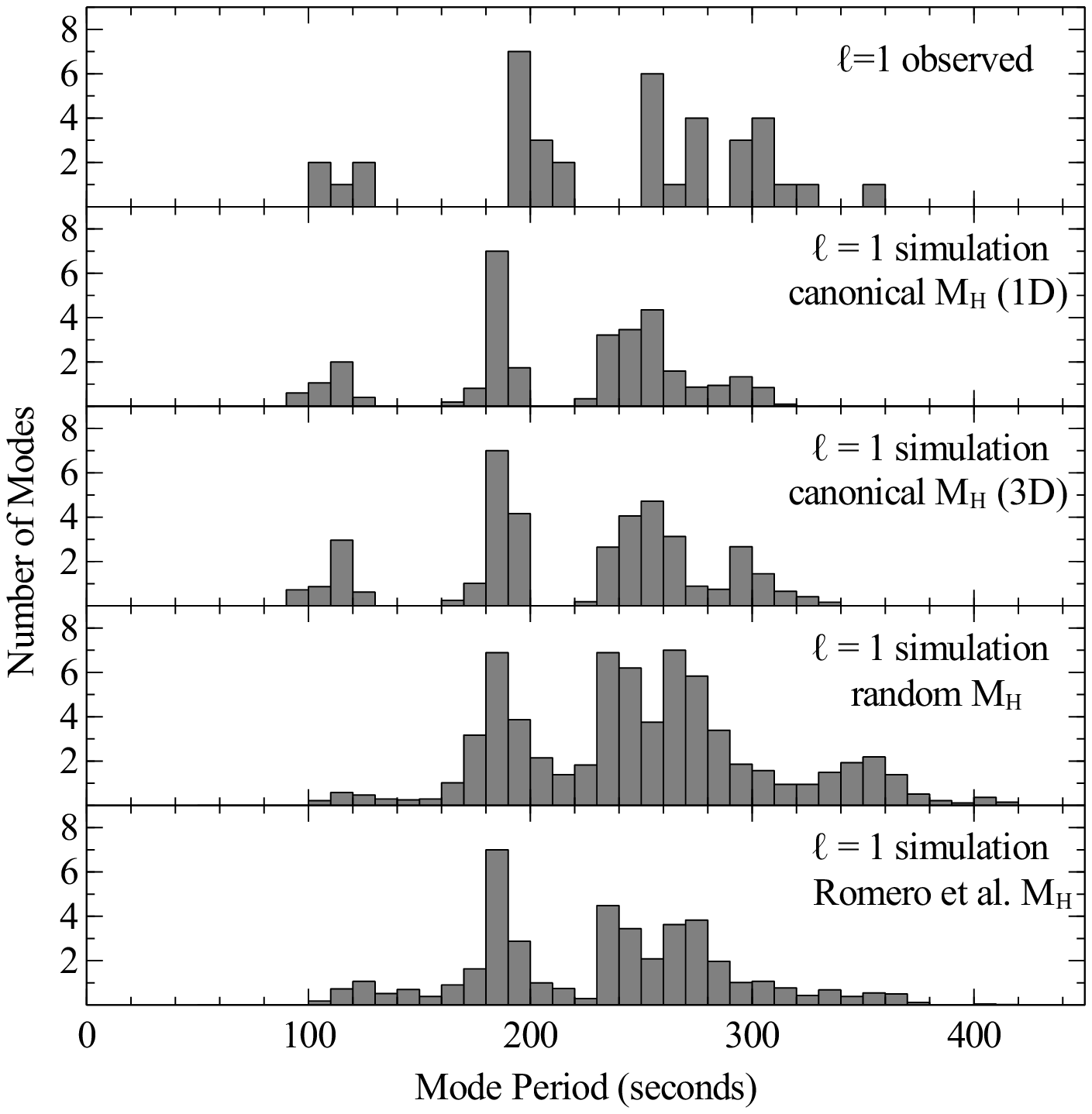}{fig:clemens_j_fig4}{The top panel shows a histogram of all the periods identified as $\ell=1$ in a sample of 16 hot ZZ Ceti stars for which common rotational splittings or other methods allow $\ell$ identification.  The two panels below show periods from a Monte Carlo simulation of periods drawn from the evolutionary models of \citet{romero2012} using spectroscopic masses and temperatures and canonical $M_{\rm H}$.  
The masses and temperatures in panel three were corrected to include the effects of 3D hydrodynamical simulations on the spectral fits, while those of panel two were not. The two lower panels are simulations in which $M_{\rm H}$ was assumed to be random, or to match the distribution that Romero found for their sample of ZZ Ceti stars.  All simulations have been normalized so that the largest bin matches the largest bin in the data.}

\acknowledgements  The authors acknowledge support from the National Science Foundation under award AST-1413001. Support for this work was also provided by NASA through Hubble Fellowship grant \#HST-HF2-51357.001-A, awarded by the Space Telescope Science Institute, which is operated by the Association of Universities for Research in Astronomy, Incorporated, under NASA contract NAS5-26555. Figure 4 was made with Veusz, a free scientific plotting package written by Jeremy Sanders. Verusz can be found at http://home.gna.org/veusz/.

\def\memsai{Mem. Soc. Astron. Ital.} 
\def\procspie{Proc. SPIE Conf. Ser.} 


\begin{thebibliography}{}

\bibitem[Bogn{\'a}r et al.(2016)]{bognar2016} Bogn{\'a}r, Z., Papar{\'o}, M., Moln{\'a}r, L., et al.\ 2016, \mnras, 461, 4059
\bibitem[Clemens(1993)]{clemens1993} Clemens, J.~C.\ 1993, Baltic Astronomy, 2, 407
\bibitem[Clemens(1994)]{clemens1994} Clemens, J.~C.\ 1994, Ph.D.~Thesis
\bibitem[Dummy et al.(2020)]{dummy} Liebert, J., Bergeron, P., \& Holberg, J.~B.\ 2005, \apjs, 156, 47 
\bibitem[Giammichele et al.(2015)]{giammichele2015} Giammichele, N., Fontaine, G., Bergeron, P., et al.\ 2015, \apj, 815, 56 
\bibitem[Gianninas et al.(2011)]{gianninas2011} Gianninas, A., Bergeron, P., \& Ruiz, M.~T.\ 2011, \apj, 743, 138
\bibitem[Hermes et al. (2016)]{hermes2016} Hermes, J. et al.\ 2016 {\it in preparation}
\bibitem[Mullally et al.(2008)]{mullally2008} Mullally, F., Winget, D.~E., Degennaro, S., et al.\ 2008, \apj, 676, 573-583
\bibitem[Robinson et al.(1995)]{robinson1995} Robinson, E.~L., Mailloux, T.~M., Zhang, E., et al.\ 1995, \apj, 438, 908
\bibitem[Romero et al.(2012)]{romero2012} Romero, A.~D., C{\'o}rsico, A.~H., Althaus, L.~G., et al.\ 2012, \mnras, 420, 1462
\bibitem[Tremblay et al.(2013)]{tremblay2013} Tremblay, P.-E., Ludwig, H.-G., Steffen, M., \& Freytag, B.\ 2013, \aap, 559, A104
\bibitem[Tremblay et al.(2016)]{tremblay2016} Tremblay, P.-E., Cummings, J., Kalirai, J.~S., et al.\ 2016, \mnras, 461, 2100
\bibitem[Yeates et al.(2005)]{yeates2005} Yeates, C.~M., Clemens, J.~C., Thompson, S.~E., \& Mullally, F.\ 2005, \apj, 635, 1239
\end{thebibliography}
\end{document}